\documentclass[12pt]{article}
\usepackage{vmargin}
\setpapersize{A4}
\setmarginsrb{2cm}{3cm}{2cm}{2cm}{0pt}{0pt}{12pt}{24pt}
\usepackage{cite} 
\usepackage{graphicx}
\usepackage[labelfont=bf,font=footnotesize]{caption}
\usepackage{authblk}
\usepackage{enumitem}
\usepackage[labelformat=simple]{subcaption}
\usepackage{floatpag}
\usepackage{amsmath}
\usepackage{amssymb}
\newtheorem{defi}{Definition}
\newtheorem{theorem}{Theorem}

\newtheorem{proposition}{Proposition}

\usepackage{longtable}
\usepackage{chngpage}
\usepackage[breaklinks]{hyperref}
\hypersetup{
	colorlinks=true,
	linkcolor=blue,
	citecolor=blue
}
\usepackage{relsize}
\usepackage[titletoc]{appendix}

\floatpagestyle{empty}
\usepackage{bm}
\usepackage{breakurl}
\DeclareMathOperator{\Tr}{Tr}

\DeclareMathOperator{\Imag}{Im}
\DeclareMathOperator{\Real}{Re}

\begin{document}

\title{Symmetric and Asymmetric Tendencies in Stable Complex Systems}
\author[1,2]{James P.L. Tan}
\affil[1]{Interdisciplinary Graduate School, Nanyang Technological University, 50 Nanyang Avenue, Block S2-B3a-01, Singapore 639798, Republic of Singapore}
\affil[2]{Complexity Institute, Nanyang Technological University, 60 Nanyang View, Singapore 639673, Republic of Singapore}
\date{}
\maketitle
\abstract{A commonly used approach to study stability in a complex system is by analyzing the Jacobian matrix at an equilibrium point of a dynamical system. The equilibrium point is stable if all eigenvalues have negative real parts. Here, by obtaining eigenvalue bounds of the Jacobian, we show that stable complex systems will favor mutualistic and competitive relationships that are asymmetrical (non-reciprocative) and trophic relationships that are symmetrical (reciprocative). Additionally, we define a measure called the interdependence diversity that quantifies how distributed the dependencies are between the dynamical variables in the system. We find that increasing interdependence diversity has a destabilizing effect on the equilibrium point, and the effect is greater for trophic relationships than for mutualistic and competitive relationships. These predictions are consistent with empirical observations in ecology. More importantly, our findings suggest stabilization algorithms that can apply very generally to a variety of complex systems. }

\newpage

Complex systems may undergo transitions between alternate stable states of contrasting behavior. Such a transition is called a critical transition or a regime shift in the literature \cite{SchefferScience}. Critical transitions are highly non-linear phenomena in that a small change in a controlling parameter such that a critical point is crossed can unexpectedly provoke a huge response (critical transition). Further away from the critical point, such a small change in the controlling parameter would only result in a comparable response without any critical transition. This non-linear response, along with the fact that critical transitions are common in nature \cite{SchefferScience,VDakosPNAS,VeraartNature,CarpenterScience,LeemputPNAS,JPLTanEPJB}, makes the study of critical transitions an important one. Critical transitions can happen as a result of instability in the stable state that the system was residing in. 

In order to determine the stability of an equilibrium point, the simplest kind of stable state, a commonly used approach in non-linear dynamics is to linearize the dynamical equations describing the system about the equilibrium point. One obtains from this linearization the $n \times n$ Jacobian matrix $\mathbf{B}$ evaluated at the equilibrium point, with real matrix elements $\{b_{ij}:i,j=1,\dots,n\}$ for a system with   dynamical variables $\mathbf{x}=(x_1,x_2,\dots,x_n)$. The matrix $\mathbf{B}$ is also known as the community matrix. The equilibrium point is stable if all real parts of the eigenvalues of $\mathbf{B}$ are negative and unstable otherwise. Henceforth in this paper, we may refer to $\mathbf{B}$ being stable or unstable when we actually mean the equilibrium point associated with $\mathbf{B}$ being stable or unstable respectively. 

The matrix element $b_{ij}$ describes the dependence of dynamical variable $x_i$ on dynamical variable $x_j$, where $i\neq j$. Conversely, $b_{ij}$ describes the dependence of $x_j$ on $x_i$. We may also refer to $b_{ij}$ as an interaction and its magnitude as its interaction strength. Here, we define the product $b_{ij}b_{ji}$ to be the relationship between $x_i$ and $x_j$. The relationship between $x_i$ and $x_j$ is mutualistic if $b_{ij}>0$ and $b_{ji}<0$, competitive if $b_{ij}<0$ and $b_{ji}<0$, and trophic if $b_{ij}b_{ji}<0$. A relationship is symmetrical when $b_{ij}$ and $b_{ji}$ are of comparable magnitudes and is asymmetrical otherwise. For example, a measure of asymmetry for mutualistic relationships is $|b_{ij}-b_{ji}|/\max (b_{ij},b_{ji})$ from Bascompte et al. \cite{BascompteScience}. The main result of this paper involves using eigenvalue bounds to show that stability in $\mathbf{B}$ favors mutualistic and competitive relationships that are asymmetrical and trophic relationships that are symmetrical. The analysis presented here stems from a rather old research question: how do the eigenvalues of $\mathbf{B}$ depend on its matrix elements?

Unfortunately, there is no exact answer to this question. An approach has been to use Random Matrix Theory (RMT), originally introduced by Wigner to study spectral properties of atomic nuclei \cite{WignerAnnMath}. RMT has since found applications in a wide variety of disciplines including number theory\cite{RudnickDuke} and neuroscience\cite{SompolinskyPhysRevLett}. In ecology, RMT was used by Robert May to study the stability of a large ecological network at an equilibrium point \cite{MayNature}. In May’s seminal work, $\mathbf{B}$ is a random matrix, with off-diagonal matrix elements being independent and identical random variables of mean zero and variance $\sigma^2$. The diagonal elements, set at -1, represent characteristic return rates for the populations of species when disturbed from equilibrium. May claimed that for large $n$, $\mathbf{B}$ is unstable when $\sigma \sqrt{n}>1$. The main criticism with May’s work is that real-world ecosystems are structured unlike the random matrix studied by May\cite{BascomptePNAS,PimmNature2,ThebaultScience}. Allesina and Tang, relying on recent advances in RMT from the mathematics literature\cite{Tao1}, recently confirmed May’s claim and further analyzed random matrices with various structures\cite{SAllesinaNature}, alleviating some of the criticisms associated with May’s work. Research in RMT has hinted that high correlation between random variables $b_{ij}$ and $b_{ji}$ in mutualistic relationships has a destabilizing effect whereas low correlation in trophic relationships has a stabilizing effect on $\mathbf{B}$\cite{AllesinaPopulationEcology,Nguyen1,TangEcolLett1}. Conjectures in RMT typically assume at the least that $n$ is large and that matrix elements or pairs of matrix elements are independently and identically distributed. The significance of the work presented here is the generality of our results: in fact we make no assumptions about $\mathbf{B}$ (besides the matrix elements being real). At the same time, we cannot obtain precise conditions for stability or instability beyond the observation that $\mathbf{B}$ will eventually become unstable if certain quantities become large enough. 

In the next few sections, we will first present the eigenvalue bounds in terms of the matrix elements and the complex parts of the eigenvalues. Then we will show that the system will become unstable when the off-diagonal sum $\chi_{\text{off}}=2\sum_i\sum_{j=i+1}b_{ij}b_{ji}$ becomes large enough. Next, we will demonstrate a stabilization algorithm on random matrices using a random strategy, a variance-minimizing strategy, and a $\chi_{\text{off}}$-minimizing strategy. This will be followed by a description of a model of $\mathbf{B}$ with ecologically motivated constraints on the interaction strengths. In that section, we also analyze the effect of dispersion in the interaction strengths on $\chi_{\text{off}}$. Finally, a discussion of the results concludes the paper. 

\subsection*{Mathematical Formulation and Eigenvalue Bounds}

We start with a dynamical system with $n$ dynamical variables described by $n$ arbitrary non-linear differential equations i.e. $\dot{x}_i=f_i(\mathbf{x})$, where $i=1,\dots ,n$ and $\mathbf{x}=(x_1,\dots,x_n)$ is a vector of dynamical variables. $\mathbf{x^*}$ is an equilibrium point if for every $i$, $f_i(\mathbf{x^*})=0$. The local stability of $\mathbf{x^*}$ may be studied by linearizing the dynamical equations about $\mathbf{x^*}$ \cite{Strogatz}. The linearization furnishes $\mathbf{B}$, the Jacobian matrix evaluated at $\mathbf{x^*}$. The matrix element $b_{ij}$, which we described as the dependence of $x_i$ on $x_j$, is the gradient of $f_i(\mathbf{x})$ along $x_j$ at $\mathbf{x^*}$ i.e. $\partial f_i/ \partial x_j(\mathbf{x^*})$. The equilibrium point is stable when any perturbation of $\mathbf{x}$ from $\mathbf{x^*}$ decays with time. Conversely, the equilibrium point is unstable when any perturbation of   from   grows with time. Stability is determined by the eigenvalues of $\mathbf{B}$. The equilibrium point is stable if the real parts of all eigenvalues are negative and is unstable otherwise. Equivalently, the equilibrium point is stable if the largest real part of all eigenvalues, which we shall refer to as the maximum real eigenvalue, is negative and unstable otherwise. Eigenvalues are the exponential decay rates of small perturbations from the equilibrium point. Thus, eigenvalues that are more negative indicate greater stability along their respective eigenvectors. Solving for the eigenvalues is equivalent to finding the roots of the characteristic polynomial $\det(\mathbf{B}-\lambda\mathbf{I})$, where $\lambda$ is an eigenvalue of $\mathbf{B}$. The eigenvalues depend on the matrix elements in a nontrivial fashion in part because there is no general algebraic expression for the roots of polynomials of the 5th degree or higher. This is the Abel-Ruffini theorem and is a well-known result from Galois theory. While others have resorted to RMT for this problem, we use an alternate approach with eigenvalue bounds to glean information about the eigenvalues’ dependence on the matrix elements. 

Given the multiset of eigenvalues of $\mathbf{B}$, $\{\lambda_i:i=1,\dots,n\}$, an upper and lower bound for the maximum real eigenvalue of $\mathbf{B}$ are respectively, 
\begin{align}
\lambda_+&=\bar{\lambda}+\sqrt{n-1}s_\lambda, \\
\lambda_-&=\bar{\lambda}+\frac{1}{\sqrt{n-1}}s_\lambda,
\end{align} 
Here, $\bar{\lambda}$ is the mean while $s_\lambda$ is the standard deviation of the real parts of all eigenvalues. The upper bound is more well-known and was probably first discovered by Laguerre but is more commonly known as Samuelson’s inequality\cite{Samuelson,STJensen1}. The lower bound is due to Brunk\cite{Brunk}. The bounds may be given by an expression in terms of $\{b_{ij}:i,j=1,\dots,n\}$ (Methods and Supplementary Information), 
\begin{align} \label{eq:egbounds}
\lambda_{\pm}=\frac{1}{n}\chi_{\text{diag}}+\frac{(n-1)^{\pm 1/2}}{\sqrt{n}} \sqrt{F_{\text{diag}} + \chi_{\text{off}} + h}
\end{align}

Here, $\chi_{\text{diag}} = \sum_i b_{ii}$ is the diagonal sum, $F_{\text{diag}}$ is a function of the diagonal elements $\{b_{ii}:i=1,\dots,n\}$ (Supplementary Information) and $\chi_{\text{off}}=2\sum_i\sum_{j=i+1}b_{ij}b_{ji}$ is the off-diagonal sum. $h=\sum_i [\Imag(\lambda_i)]^2$ is a non-negative number that is positive when the imaginary components are non-zero and zero otherwise. $n$ is kept constant throughout our analysis. The mean of the eigenvalues is controlled by the diagonal elements i.e. $\bar{\lambda}=\chi_{\text{diag}}/n$. Hence, the eigenvalues’ dependence on the diagonal elements is more straightforward and is generally of less interest than the off-diagonal elements. It follows from the bounds that $\lambda_+<0$ is a sufficient condition for stability while $\lambda_->0$ is a sufficient condition for instability. 

From Equation \ref{eq:egbounds}, we may then draw some conclusions on two cases: (i) the eigenvalues are real numbers, i.e. $h=0$ (ii) the eigenvalues are complex numbers, i.e. $h\geq 0$. For the first case, it is always possible to achieve stability or instability by decreasing or increasing $\chi_{\text{off}}$ respectively. The first case shall not be analyzed further because of the strong assumption that all eigenvalues are real. Instead, we consider the second case, which is more general. For the second case, the upper bound becomes less useful but not the lower bound because $h\geq 0$; we can still always increase $\chi_{\text{off}}$ enough such that $\mathbf{B}$ becomes unstable. Therefore, while it is still necessary to keep $\chi_{\text{off}}$ small enough for stability, keeping $\chi_{\text{off}}$ small alone does not guarantee stability because of $h$. For this reason, other contributing factors would still need to be considered in order to form a complete picture of how stability arises in $\mathbf{B}$. For example, ecologists have been obsessed with the nestedness, a persistent structural property observed in mutualistic networks\cite{BascomptePNAS,Staniczenko,Montoya,Suweis} In a nested architecture of a bipartite network, a more specialist species (defined as having fewer mutualistic interactions) would only interact with a proper subset of mutualistic partners of the more generalist species (defined as having more mutualistic interactions). However, there remains some controversy over how important nestedness is to the stability of mutualistic ecological networks\cite{Staniczenko}. Instead of delving into the details of specific structural properties, we will focus our efforts on minimizing $\chi_{\text{off}}$ instead. $\chi_{\text{off}}$ is the sum of all relationships in $\mathbf{B}$. A natural way to minimize $\chi_{\text{off}}$ is to make both interaction strengths in mutualistic relationships very weak. However, mutualistic relationships are pervasive in nature. Therefore, we need to constrain the interaction strengths in $\mathbf{B}$ so that minimization of $\chi_{\text{off}}$ will not render both interaction strengths in $\mathbf{B}$ negligible. Then, minimization of $\chi_{\text{off}}$ will require mutualistic relationships to be asymmetric in order to minimize each summand, $b_{ij}b_{ji}$ i.e. one large and one small interaction strength. In the section on interdependence diversity and symmetric correlation, we will constrain the interaction strengths in $\mathbf{B}$ from an ecological standpoint and demonstrate that minimization of $\chi_{\text{off}}$ will require mutualistic relationships to be asymmetrical and trophic relationships to be symmetrical. 

\subsection*{Pruning Random Matrices for Stability}

The results presented thus far suggests that minimization of $\chi_{\text{off}}$ might provide an efficient route to stabilize an equilibrium point. We employ a simple algorithm on a well-known example, the random matrix studied by May \cite{MayNature}. For this example, consider the situation where $\mathbf{M}$ is a random matrix and its diagonal elements are set at $-d$ while the off-diagonal elements are independently and identically distributed random variables of mean zero and variance $\sigma^2$. According to RMT, for large $n$, the eigenvalues of $\mathbf{M}$ are contained in a circle of radius $\sigma \sqrt{n}$ centered at $(-d,0)$ on the complex plane\cite{SAllesinaNature}. For this example, we use $n=20$, $d=2$, a standard normal distribution for the off-diagonal elements and a modification factor $g$ that we shall introduce in the description of the algorithm. 
\begin{figure}[h!]
\centering
\centerline{\includegraphics[scale=0.4]{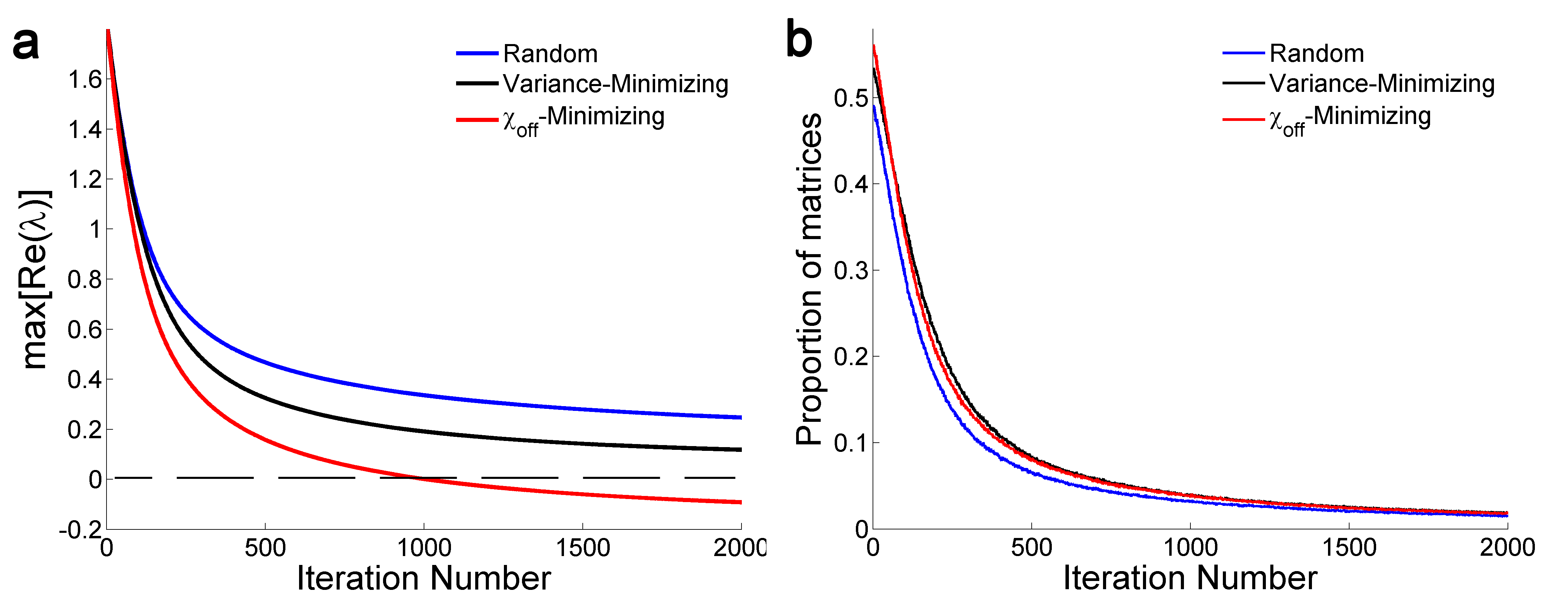}}
\caption{\textbf{Pruning unstable random matrices.} Results of the stabilization algorithm employed on 50,000 unstable $20 \times 20$ random matrices for the three different stabilization strategies (random, variance-minimizing and $\chi_{\text{off}}$-minimizing) described in the main text. Data points at each iteration indicate the sample average over the 50,000 simulations. Standard error of the mean estimates are on the order of $10^{-3}$ for both figures. (a) The maximum real eigenvalue at the end of each iteration. (b) The proportion of all matrices with a decreased maximum real eigenvalue from the previous iteration number. }
 \label{fig:StabilizeMatrices}
\end{figure}
The algorithm consists following steps: (1) initialize a random matrix $\mathbf{M}$, (2), calculate the eigenvalues of $\mathbf{M}$, (3) choose an off-diagonal matrix element $b_{ij}$ randomly, (4) if $b_{ij}b_{ji}<0$, multiply $b_{ij}$ by a factor $g$, else if $b_{ij}b_{ji}>0$, divide $b_{ij}$ by a factor $g$, (5) calculate the new eigenvalues of $\mathbf{M}$ after the modification, and (6) if the maximum real eigenvalue of   after the modification is larger than before the modification, revert to step (2) using $\mathbf{M}$ before the modification; if the maximum real eigenvalue of $\mathbf{M}$ after the modification is smaller than before the modification instead, revert to step (2) using $\mathbf{M}$ after the modification. This counts as one iteration. 

This algorithm employs a $\chi_{\text{off}}$-minimizing strategy due to step (4). We compare this algorithm using the $\chi_{\text{off}}$-minimizing strategy against the same algorithm using a random strategy and a variance-minimizing strategy. In the random strategy, step (4) is replaced by the following step instead: (4) $b_{ij}$ is randomly chosen to be multiplied or divided by $g$ with probability ½. In the variance-minimizing strategy, step (4) is replaced by the following step instead: (4) $b_{ij}$ is divided by $g$. We compare the three strategies using 50,000 random matrices over 2,000 iterations. The results are shown in Figure \ref{fig:StabilizeMatrices}. The $\chi_{\text{off}}$-minimizing strategy clearly outperforms the other two strategies. Of course, if we were to accept every modification without checking if it reduces the maximum real eigenvalue at every iteration, then the variance-minimizing strategy will eventually reduce all eigenvalues to $-d$. However, there are two reasons why such an algorithm might be undesirable: (i) the maximum real eigenvalue may at times increase with iteration number, and (ii) the eventual interaction strengths are small unlike the original algorithm with the $\chi_{\text{off}}$-minimizing strategy which allows for larger eventual interaction strengths. Sensitivity analysis of the parameter $g$ reveals that the  $\chi_{\text{off}}$-minimizing strategy still outperforms the other two strategies for the various values of $g$ tested (Supplementary Information). Statistics of the matrices at the end of the iterations and figures displaying matrices after implementations of the stabilization algorithm can be found in the Supplementary Information. 

\subsection*{Interdependence Diversity and the Symmetric Correlation}

Clearly, it is not realistic to stabilize $\mathbf{B}$ by rendering the interaction strengths in $\mathbf{B}$ negligible since interactions are ubiquitous in nature. Therefore, we now consider ecologically motivated constraints of the interaction strengths in $\mathbf{B}$. To do this, we first consider two sets of similar variables, $y=\{y_k:k=1,\dots,m\} \subset x$ and $z=\{z_l:l=1,\dots,m\}\subset x$, where $x$ is the set of all variables and $y\cap z = \varnothing$. These two sets of variables are so defined to delineate interactions of a particular type between variables from the two sets. For example, if the interaction types are consumption and pollination, then the variables in $y$ could represent populations of pollinators while the variables in $z$ will represent populations of plants. We now formulate equations of constraints that allow variables in $y$ to depend on various weighted combinations of the variables in $z$ and vice versa. For notational convenience, let us denote $Y_k(\mathbf{x})$ to be $Y_k(\mathbf{x})=\dot{y}_k$ and $Z_l(\mathbf{x})$ to be $Z_l(\mathbf{x})=\dot{z}_l$ for all $k$ and $l$. Then, we may find in $\mathbf{B}$ the matrix elements $\partial Y_k/ \partial z_l(\mathbf{x^*})=d_{kl}\alpha_k$, which is the dependence of species $y_k$ on species $z_l$, and $\partial Z_l / \partial y_k (\mathbf{x^*})=e_{lk}\beta_l$, which is the dependence of species $z_l$ on species $y_k$, for all $k$ and $l$. Here, $d_{kl}$ and $e_{lk}$ are weights such that $\sum_k d_{kl}=1$, $\sum_l d_{kl}=1$, $\sum_k e_{lk}=1$, $\sum_l e_{lk}=1$, and $0<d_{kl},e_{lk}<1$. $\alpha_k$ and $\beta_l$ are real numbers and because of the bounded weights, their absolute values are the maximum interaction strengths possible for the respective interactions (e.g. consumption and pollination) and species ($y_k$ and $z_l$) they pertain to. These constraints on the interaction strengths can be motivated by the constant interacting effort hypothesis which states that interaction strengths should be stronger, on average, for species interacting with a smaller number of resource species \cite{SuweisOikos}. This hypothesis was postulated based on the fact that there is a limited amount of time a species has to interact with other species. Hence, if the interaction strengths are to be proportional to the time spent on interacting with other species, and that there is a fixed amount of time for a type of interaction, then we arrive at the constraints $\sum_l d_{kl}=1$ and $\sum_k e_{lk}=1$ for all $k$ and $l$. While a species can spend a different amount of time for a type of interaction at the expense of other types of interactions, it is sufficient to fix the total amount of time spent on a type of interaction for the purpose of analyzing the effect of dispersion in the weight distributions on the off-diagonal sum. Additionally, we might also require that the dependencies on any species ($y_k$ or $z_l$) be constrained ($\sum_l e_{lk}=1$ and $\sum_k d_{kl}=1$ respectively) although these additional constraints do not affect the conclusions that are about to follow (Theorem S1). Finally, differences in $\alpha_k$ for all $k$ and differences in $\beta_l$ for all $l$ lead to biases in weight distribution amongst the variables in $y$ and $z$ when minimizing $\chi_{\text{off}}$. Since our goal is to evaluate the effect of the dispersion in the weights on the off-diagonal sum, we may simplify the problem by assuming that $\alpha_k=\alpha$ for all $k$ and $\beta_l=\beta$ for all $l$. 

The relationship between any variable $y$ and any variable in $z$ is either mutualistic or competitive if $\alpha\beta >0$ and trophic if $\alpha\beta <0$. If we arrange the weights $d_{kl}$ into an $m \times m$ matrix $\mathbf{D}$ such that $d_{kl}$ is a matrix element of $\mathbf{D}$, then $\alpha \mathbf{D}$ is a submatrix of $\mathbf{B}$. Similarly, $e_{lk}$ is a matrix element of $\mathbf{E}$ and $\beta \mathbf{E}$ is a submatrix of $\mathbf{B}$. This convenience allows us to define a quantity $C$ that we shall call the symmetric correlation, 
\begin{align}
C=m^{-1}\sum_k\sum_l d_{kl} e_{lk} = m^{-1}\Tr(\mathbf{D}\mathbf{E}). 
\end{align}
The symmetric correlation is bounded $0<C<1$ (Theorem S2) and contains information about both the variance of the weight distributions and the correlation between the matrix elements of $\mathbf{D}$ and the corresponding matrix elements of $\mathbf{E}^T$. For example when $C=1$, all weights are either zeros or ones and they fulfill $d_{kl}=e_{lk}$ for all $l$ and $k$ whereas when $C=0$, all weights fulfill $d_{kl}e_{lk}=0$ for all $l$ and $k$. When the variance of the weight elements in $\mathbf{D}$ and the variance of the weight elements in $\mathbf{E}$ are fixed, we may use $C$ as a relative measure of symmetry and correlation between the matrix elements in $\mathbf{D}$ and the corresponding matrix elements in $\mathbf{E}^T$ (Methods). We find that $\chi_{\text{off}}$ contains the summand $m\alpha\beta C$ i.e. $\chi_{\text{off}}=2m\alpha\beta C + \dots$, and that the weight elements $d_{kl}$ and $e_{lk}$ for all $k$ and $l$ are contained exclusively in the summand $2m\alpha\beta C$ of $\chi_{\text{off}}$. Thus, minimizing $\chi_{\text{off}}$ for the relationship $\alpha \beta$ requires minimizing $C$ for mutualistic and competitive relationships and maximizing $C$ for trophic relationships i.e. mutualistic and competitive relationships will be asymmetric whereas trophic relationships will be symmetric (specifically, the weights associated with trophic relationships will be symmetric). 
\begin{figure}[h!] 
\centering
\centerline{\includegraphics[scale=0.7]{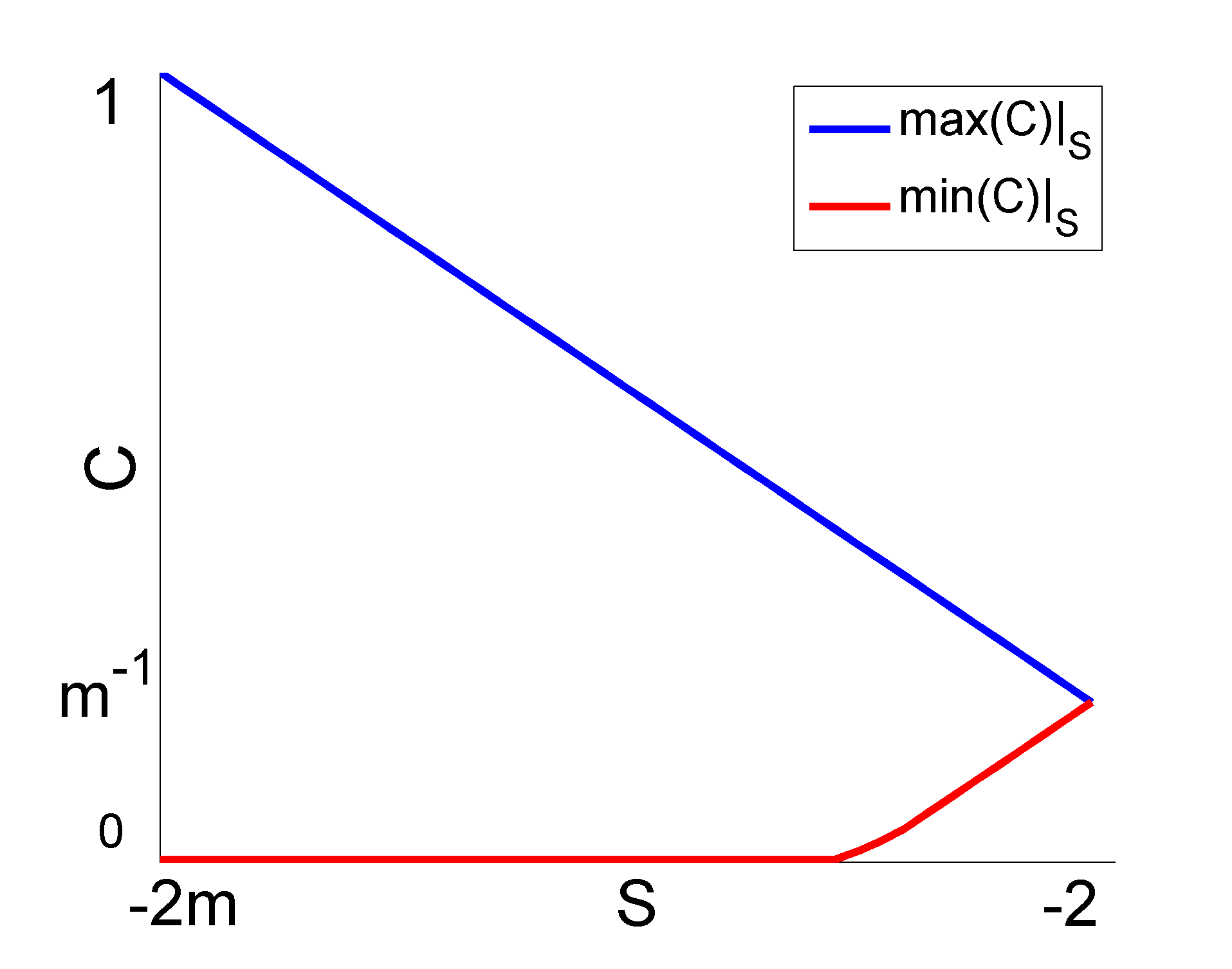}}
\caption{\textbf{Symmetric correlation and interdependence diversity.} This graph shows the boundaries of values possible for $C$ and $S$. The blue line is the maximum $C$ attainable under fixed $S$. The red line is the minimum $C$ attainable under fixed $S$. Both line plots are calculated by numerical optimization techniques with $m=$ (Methods). An analytical calculation for $m \geq 1$ is provided in the Supplementary Information. As a further note, the set of $C$ for a fixed $S$ is not necessarily continuous within the boundaries (e.g. at $S=-2m$). }
\label{fig:InterdependencyCurve}
\end{figure}

Next, we define the interdependence diversity, 
\begin{align}
S=-\sum_{k=1}^{m} \sum_{l=1}^{m} {d_{kl}}^2 + {e_{lk}}^2,
\end{align}
a measure of the diversity of dependencies among the $y$ and $z$ variables. $S$ is simply the sum of all the squared weight elements. Furthermore, due to the weight constraints, $S$ is bounded $-2m<S\leq -2$. When $S=-2m$ at minimum interdependence diversity, then all weights are either zeros or ones. When $S=-2$ at maximum interdependence diversity, then all weights are equal to $1/m$. The interdependence diversity defined here is similar to the Herfindahl index in economics \cite{Herfindahl} or the Simpson index in ecology \cite{Simpson}. We denote $\max(C)|_S$ and $\min(C)|_S$ to be respectively the maximum and minimum $C$ under any variation of weights and under a fixed $S$ (without violating the weight constraints). The relationships of $\max(C)|_S$ and $\min(C)|_S$ with $S$ are shown using numerical calculations in Figure \ref{fig:InterdependencyCurve}. Analytical calculations can be found in the Supplementary Information. Minimizing $\chi_{\text{off}}$ means that mutualistic relationships will reside on the $\min(C)|_S$ curve while trophic relationships will reside on the $\max(C)|_S$ curve. Both $\max(C)|_S$ and $\min(C)|_S$ are monotonically decreasing and increasing functions of $S$ respectively (Figure \ref{fig:InterdependencyCurve}). Hence, the capacity of $\mathbf{B}$ to minimize $\chi_{\text{off}}$ decreases with increasing interdependence diversity for both mutualistic, competitive and trophic relationships. Additionally, because $\max(C)|_S$ is more adversely affected than $\min(C)|_S$ with increasing interdependence diversity (for $m>2$, Figure \ref{fig:InterdependencyCurve}, this effect is more pronounced in trophic relationships than mutualistic and competitive relationships. Essentially, trophic relationships are more affected than mutualistic relationships with increasing interdependence diversity because there exists many more possibilities in the network to minimize $C$. Only when the network is fully connected with equally weighted one-directional links does $\min(C)|_S$ start increasing with $S$ (Proposition S2). 

\subsection*{Discussion}
In this work, we have derived eigenvalue bounds for the maximum real eigenvalue of $\mathbf{B}$ in terms of the matrix elements and the complex eigenvalues. From these bounds it follows that a necessary condition for stability is that $\chi_{\text{off}}$ is small, for $\chi_{\text{off}}$ can always be increased enough such that $\mathbf{B}$ will become unstable. The generality of this result and subsequent calculations allows us to consider different types of interactions in concert, something that was limited in previous studies with RMT due to assumptions on matrix elements being independently and identically distributed. Additionally, we show that two observations, increasing interdependence diversity causing decreasing $\chi_{\text{off}}$ and this decrease in $\chi_{\text{off}}$ being more pronounced for trophic than mutualistic and competitive relationships, can both be explained as a result of $\mathbf{B}$ losing its capacity to accommodate symmetric and asymmetric relationships. 

In the course of implementing the stabilization algorithm, the maximum real eigenvalue will be a monotonically decreasing function of iteration number. From the eigenvalue bounds, we generally expect $h+\chi_{\text{off}}$ to also decrease with iteration number. Indeed, statistics of the matrices after 2,000 iterations of the stabilization algorithm reveal that the average change in $h+\chi_{\text{off}}$ is negative for all three strategies, with the effect of decreasing $\lambda_{\pm}$ from the initial random matrices (Table S1). In particular, $h$ and the standard deviation of the off-diagonal elements increase for the $\chi_{\text{off}}$-minimizing strategy, with the mean of the off-diagonal elements remaining constant. This suggests that while there is a certain risk in increasing $h$ when increasing the interaction strengths, it is possible that the increase in $h$ can be mitigated and overcome by a larger decrease in $\chi_{\text{off}}$ such that the system can be stabilized with increasing interaction strengths. 

The interplay between $h$ and $\chi_{\text{off}}$ is an important factor to consider when minimizing $\chi_{\text{off}}$ to stabilize $\mathbf{B}$. We have shown, under a general framework of ecologically motivated constraints, that minimization of $\chi_{\text{off}}$ will result in trophic relationships being more adversely affected than mutualistic and competitive relationships with increasing interdependence diversity. The validity of this result for lowering the eigenvalue bounds of an ecological community will depend on whether a not minimization of $\chi_{\text{off}}$ will necessarily give rise to an increase of $h$ larger than the decrease in $\chi_{\text{off}}$ in every minimization scenario under the general framework of constraints employed. Empirical observations in the ecology literature suggest that this may not the case for most communities. Our result is consistent with empirical observations if we allow the interdependence diversity defined here to be used as a proxy for connectance, a measure that is well known in the ecology community. The connectance is the proportion of non-zero dependencies in $\mathbf{B}$. Hence, we generally expect an increasing connectance to also result in an increasing interdependence diversity as the number of interactions increases and as the interaction strengths become more distributed among the dynamical variables. Thébault and Fontaine found trophic networks to have a lower connectance than mutualistic networks in a meta-analysis of real-world pollination (mutualistic) and herbivory (trophic) networks while controlling for $n$ \cite{ThebaultScience}. Therefore, our derived result that increasing interdependence diversity having a destabilizing effect being more pronounced in trophic relationships than mutualistic and competitive relationships could provide a plausible theoretical explanation for this empirical observation.

The prediction that mutualistic and competitive relationships are symmetric whereas trophic relationships are asymmetric also agrees with empirical observations. It has been known for some time that mutualistic ecological networks like plant-pollinator networks consist of highly asymmetric interactions between plant and pollinator \cite{BascompteScience, BascomptePNAS, Jordano}. For example, the manduvi tree relies almost exclusively on the toco toucan for seed dispersal, but the toco toucan is not limited to the manduvi tree’s fruits in its diet \cite{BascompteAnnRev1}. Overall, consistency of our calculations with empirical observations demonstrates our approach to be promising for further investigations of stability in $\mathbf{B}$.

Our results highlight the importance of asymmetry in mutualistic and competitive relationships, and of symmetry in trophic relationships to the stability of a complex system. Identifying and understanding the contributing factors to stability can be used to help design algorithms to stabilize real-world systems on the verge of critical transitions. For example, the stabilization algorithm described in this paper could be a starting point for future investigations into the stabilization of real-world systems. In a successful realization of such an algorithm, critical slowing down signals could be used to measure the change in stability at every iteration (step (5) of the algorithm). Critical slowing down signals are statistical signals that can be used to detect if a stable state is becoming more unstable. These signals have been detected in a wide variety of real-world systems\cite{SchefferScience}. They are based on the premise of a slower return rate to the stable state after a perturbation as the stable state becomes more unstable\cite{SchefferNature}. While there have been ample studies on the detection of critical slowing down signals, more research needs to be conducted on the stabilization of potentially unstable stable states. 

Stabilization is one way to deal with critical transitions. A recent attempt at this problem involves smoothening the non-linearity of a critical transition\cite{Martin}. Network properties not covered in this work can also be very important in dealing with instability. For a formerly stable equilibrium point, initial instability occurs when the maximum real eigenvalue goes above zero. The eigenvector(s) of the maximum real eigenvalue determine the initial directions of instability and which variables will be initially affected by this instability. As the system transitions away from the previously stable equilibrium point, more and more variables might be affected depending on their dependence on the initially and subsequently affected variables in what is known as a cascade of failures. Whether a not such an initial instability will eventually lead to system-wide instability depends on a multitude of factors including the structure of the network connecting these variables and how the system responds to this initial instability. For example, in a load bearing network with a heterogeneous degree distribution, the failure of a single node with a large number of dependencies can cause a large cascade of failures\cite{Motter}. In ecological mutualistic networks, the right and left leading eigenvectors not only determine the species affected by perturbations to the system and the size of these perturbations, they also positively correlate with a few network properties like the degree centrality and the page-rank centrality \cite{SuweisNatComm}. The effect of initial instability or failure on the whole system is a topic of great interest in network science\cite{Motter,Buldyrev,Watts}. While there remains a host of factors that ultimately determine stability in a complex system, the generality of our results suggests that asymmetry in mutualistic relationships and symmetry in trophic relationships should be universally observed and not restricted to ecology. 

\subsection*{Methods}
\paragraph{Derivation of eigenvalue bounds.} The polynomial equation is $\det(\mathbf{B}-\lambda\mathbf{I})=\lambda^n + c_1 \lambda^{n-1} + c_2 \lambda^{n-2} + \dots$. We may express both bounds in terms of $c_1$, $c_2$ and $h$ using Viète’s formulas and the complex conjugate root theorem. The relation between the matrix elements and the coefficients $c_1$ and $c_2$ can be found by expanding the Leibniz formula for matrix determinants. This gives us the bounds in terms of the matrix elements of $\mathbf{B}$. A more detailed derivation may be found in the Supplementary Information. 

\paragraph{Obtaining the numerical results of Figure 2.} To obtain $\max(C)|_S$, we (1) construct $5 \times 5$ matrices $\mathbf{D}_{\text{max}}$ and $\mathbf{E}_{\text{max}}$ at $\max(C)|_S$ when $S$ is at the minimum of $-2m$; $\mathbf{D}_{\text{max}}$ and $\mathbf{E}_{\text{max}}$ are initial starting points for a nonlinear constrained optimization (maximization) algorithm implemented in MATLAB (\textit{fmincon} function with \textit{sqp} algorithm where the constraints for the optimization problem are the weight constraints $0<d_{kl},e_{lk}<1$, $\sum_k d_{kl}=1$, $\sum_l d_{kl}=1$, $\sum_k e_{lk}=1$, and $\sum_l e_{lk}=1$, and the interdependence diversity constraint $S=\sum_{k,l} {d_{kl}}^2 + {e_{lk}}^2 = -2m$, while the objective function is $C$), (2) carry out the optimization for the starting point and constraints, and (3) use the solution as the new weight matrices for the starting point of the next optimization where the interdependence diversity is fixed at a positive increment $\varepsilon = 0.001$ from the previous optimization. Steps (2) and (3) are repeated until the maximum interdependence diversity is reached at $-2$. To obtain $\min(C)|_S$, we use the same steps, replacing the initial starting point with $\mathbf{D}_{\text{min}}$ and $\mathbf{E}_{\text{min}}$ at $\min(C)|_S$ when $S=-2m$ and using the same optimization algorithm but with minimization instead. 

\paragraph{The symmetric correlation as a relative measure of symmetry and correlation. } Let $D=(d_{1,1},d_{1,2},d_{1,3},\dots )$ represent a sequence of the matrix elements of $\mathbf{D}$ and $E^T=(e_{1,1},e_{2,1},e_{3,1},\dots)$ represent the corresponding sequence of the matrix elements of $\mathbf{E}^T$. A measure of correlation between $D$ and $E^T$ is the Pearson’s correlation coefficient estimate
\begin{align}
r_{D,E^T}=\frac{\overline{DE^T}-\overline{D}\overline{E^T}}{s_D s_E}, 
\end{align}
where $\overline{DE^T}$, $\overline{D}$ and $\overline{E^T}$ are sample means, $s_D$ is the standard deviation of $D$ and $s_E=s_{E^T}$ is the standard deviation of $E^T$. Because of the weight constraints, $\overline{D}=\overline{E^T}=1/m$ are constants independent of the weight distribution of $\mathbf{D}$ and $\mathbf{E}^T$. Also, $\overline{DE^T}=m^{-2}\mathbf{D}\mathbf{E}$. Hence, the symmetric correlation $C=m^{-1}\mathbf{D}\mathbf{E}$ may be used as a relative measure of symmetry or correlation between matrix elements of $\mathbf{D}$ and the corresponding matrix elements of $\mathbf{E}^T$ when $s_D$ and $s_E$ are fixed. 

\subsection*{Acknowledgements}
The author would like to thank Siew Ann Cheong and Lock Yue Chew for discussions on this project. Additionally, he would also like to thank three anonymous reviewers for their helpful suggestions and comments. In particular, the author is grateful for the suggestion of the constant interacting effort hypothesis as a motivation for introducing the constraints. 

\newpage
\section*{Supplementary Information}
\subsection*{Derivation of the eigenvalue bounds}
Given the multiset of eigenvalues $\{\lambda_i:i=1,\dots , n\}$ of the matrix $\mathbf{B}$ whose matrix elements are real, the eigenvalue bounds for the maximum real part of all eigenvalues are 
\begin{align}
\lambda_{\pm} = \bar{\lambda} + (n-1)^{\pm 1/2}s_{\lambda}
\end{align}
, where $\bar{\lambda}$ is the mean while $s_\lambda$ is the standard deviation of the real parts of all eigenvalues. Here, $\bar{\lambda} = \sum_i \Real(\lambda_i)/n$ and 
\begin{align}
s_\lambda &=\sqrt{\frac{1}{n} \sum_i {(\Real(\lambda_i) - \bar{\lambda})}^2} \\
& =\sqrt{-\bar{\lambda}^2+\frac{1}{n} \sum_i {[\Real(\lambda_i)]}^2} 
\end{align}

If we denote the polynomial equation as $\text{det}(\lambda\mathbf{I}-\mathbf{B}) =\lambda^n+c_1\lambda^{n-1}+c_2\lambda^{n-2}+\dots$, then we find using Vi\`{e}te's formulas and the complex conjugate root theorem that $c_1 = -\sum_i \lambda_i = -\sum_i \Real(\lambda_i)$ and $c_2 = \sum_i \sum_{j=i+1} \Real(\lambda_i) \Real(\lambda_j) + h/2$, where $h=\sum_i [\Imag(\lambda_i)]^2$. Then $\lambda_{\pm}$ becomes
\begin{align}
\lambda_{\pm} = -\frac{c_1}{n} + \frac{(n-1)^{1/2\pm 1/2}}{n} \sqrt{{c_1}^2 + \frac{2n}{n-1}(h/2-c_2)}. 
\end{align}
Expanding the Leibniz formula for determinants gives us $c_1 = -\sum_i b_{ii}$ and $c_2 = \sum_i\sum_{j=i+1}b_{ii}b_{jj}- b_{ij}b_{ji}$. Hence, 
\begin{align} \label{eq:bounds}
\lambda_{\pm} &= \frac{1}{n}\chi_{\text{diag}} + \frac{(n-1)^{\pm 1/2}}{\sqrt{n}} \sqrt{F_{\text{diag}} + \chi_{\text{off}} + h}, 
\end{align}
where $\chi_{\text{diag}} = \sum_i b_{ii}$, $\chi_{\text{off}} = \sum_i \sum_{j=i+1} b_{ij}b_{ji}$ and 
\begin{align}
F_{\text{diag}} = \frac{n-1}{n}\sum_i {b_{ii}}^2 - \frac{2}{n} \sum_i \sum_{j=i+1} b_{ii} b_{jj}. 
\end{align}

\subsection*{Supplementary results for the stabilization algorithm}

\subsubsection*{Sensitivity analysis of the $g$ parameter in the stabilization algorithm}
\begin{figure*}[h!]
\centering
\centerline{\includegraphics[scale=0.7]{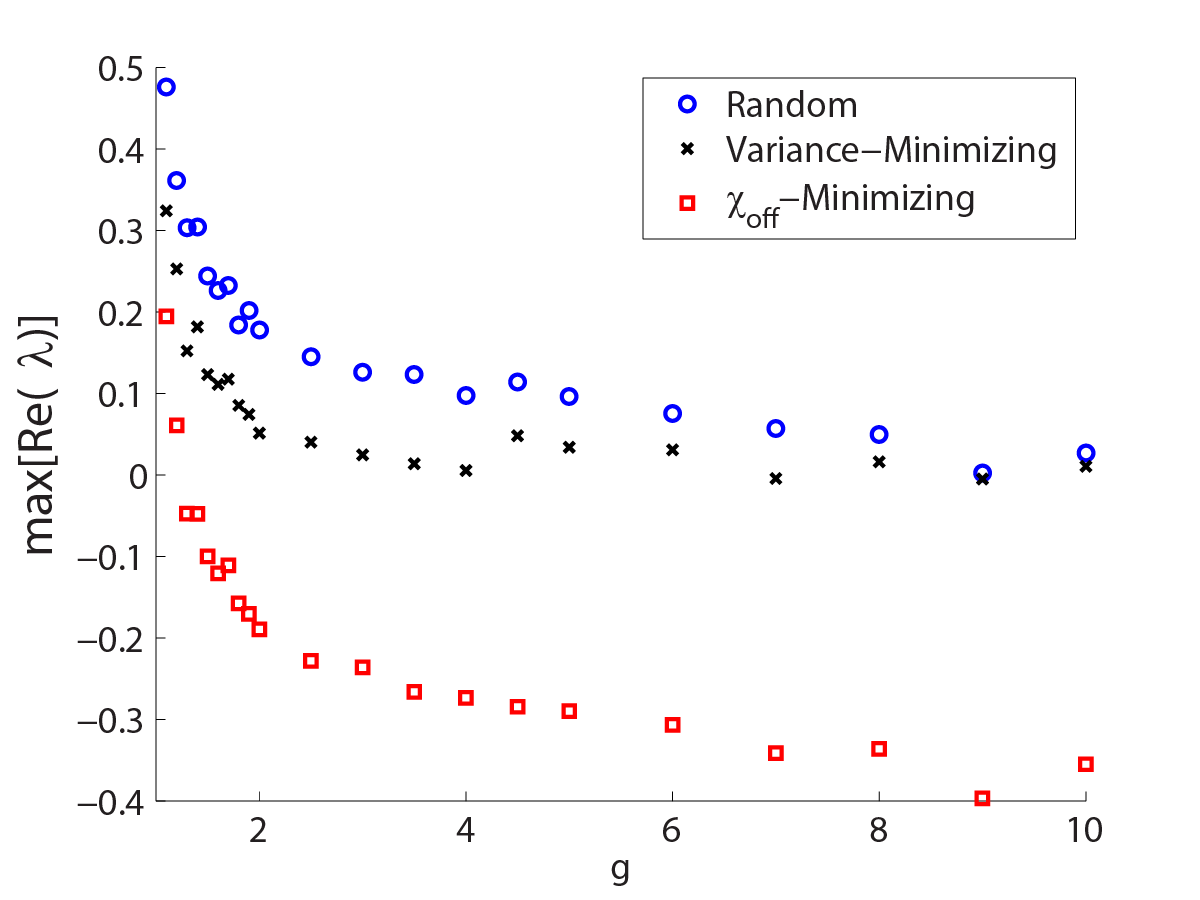}}
\caption{The average of the maximum real eigenvalues belonging to 1,000 random matrices at the end of 2,000 iterations is plotted against $g$ for the three different strategies described in the main text. Error bars are the 95\% confidence intervals for the population mean. } 
\label{fig:SensitivityAnalysisG}
\end{figure*}

The parameter $-d$ translates the eigenvalues and the parameter $\sigma$ scales the eigenvalues. Hence these two parameters are unimportant for the sensitivity analysis. The sensitivity analysis is conducted for the $g$ parameter instead, where $g>1$. The algorithm is performed on 1,000 random matrices up to an iteration length of 2,000 for various values of $g$ from 1.1 to 10. The parameters used are $n=20$, $d=2$, and random variables drawn from a standard normal distribution. Results of the sensitivity analysis are shown in Figure \ref{fig:SensitivityAnalysisG} and do not indicate that the $\chi_{\text{off}}$-minimizing strategy is worse performing than the other two strategies for any value of $g$ tested. 

\newpage
\begin{figure*}[t!]
\centering
\centerline{\includegraphics[scale=0.43]{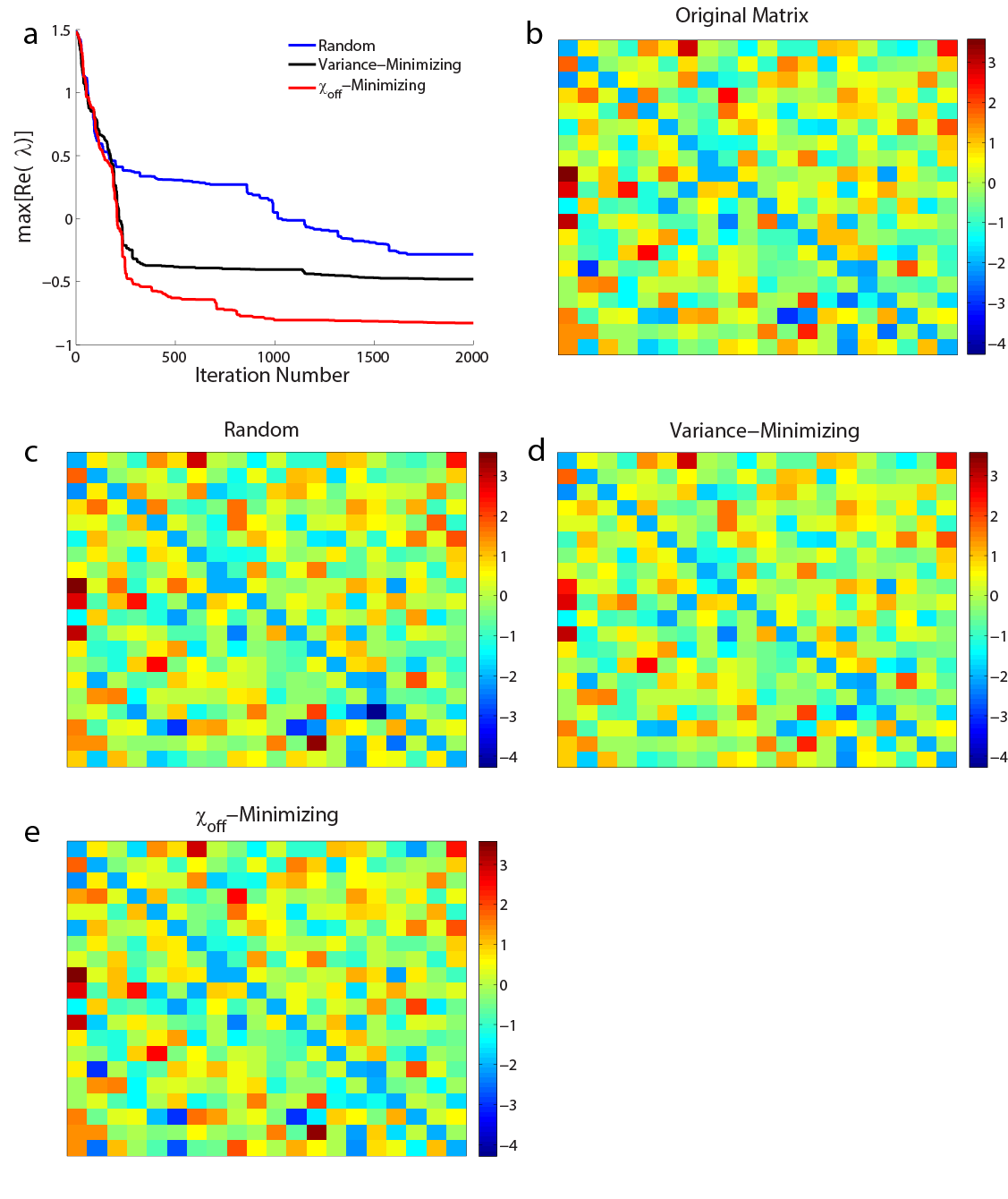}}
\caption{Heatmaps of the matrices after implementations of the stabilization algorithm with the various strategies on an initially unstable matrix. The initial $20 \times 20$ matrix is generated with diagonal elements set at -2 and off-diagonal elements drawn from a standard normal distribution. The modification factor is $g=3/2$. (a) The maximum real eigenvalue at the end of each iteration. (b) The initial matrix before stabilization. (c), (d) and (e) The matrix after 2,000 iterations with the random, variance-minimizing and $\chi_\text{off}$-minimizing strategies respectively. } 
\label{fig:Matrices}
\end{figure*}

\begin{table*}[h!]
\caption{Table of statistics from before and after 50,000 implementations of the stabilization algorithm with the three different strategies. The initial matrices before stabilization are random matrices generated as per the main text. The iteration length is 2,000 and the modification factor is $g=3/2$. Statistics shown are population means of several properties of the matrices. Population mean estimates with 99\% confidence intervals are indicated where appropriate. $\chi_{\text{off}}=2\sum_i \sum_{j=i+1} b_{ij}$ is the off-diagonal sum, $h=\sum_i[\Imag (\lambda_i)]^2$ is the sum of squared imaginary eigenvalues, $\Delta(h+\chi_{\text{off}})$ is the change in $h+\chi_{\text{off}}$ from before the stabilization, $\mu_{\text{off}}$ is the mean of the off-diagonal elements, $\sigma_{\text{off}}$ is the standard deviation of the off-diagonal elements, $\sum_i \sum_{j\neq i} |b_{ij}|H(b_{ij}b_{ji})$ (with $H(b_{ij}b_{ji})$ the Heaviside step function) is the total interaction strength of mutualistic and competitive relationships, and $\sum_i \sum_{j \neq i} |b_{ij}|H(-b_{ij}b_{ji})$ is the total interaction strength of trophic relationships. }
\centering
\begin{tabular}{l | c | c | c | c} 
 & Initial Matrix & Variance-Minimizing & $\chi_{\text{off}}$-Minimizing & Random \\ \hline 
$\chi_{\text{off}}$ & 0 & (-9.6, -8.5) & (-62.4, -60.5) & (-14.0, -12.6) \\ 
$h$ & (96.2, 96.6) & (80.4, 80.7) & (126.5, 127.1) & (96.9, 97.3) \\ 
$\Delta(h+\chi_{\text{off}})$ & - & (-24.9, -24.6) & (-31.1, -30.7) & (-12.6, -12.4) \\ 
$\mu_{\text{off}}$ & 0 & (-4.5E-4, 6.4E-4) & (-4.0E-4, 7.4E-4) & (-4.2E-4, 7.7E-4) \\
$\sigma_{\text{off}}$ & 1 & (0.928, 0.929) & (1.036, 1.037) & (1.016, 1.017) \\ 
$\sum_i \sum_{j \neq i} |b_{ij}| H(b_{ij}b_{ji})$ & 151.6 & (135.1, 135.4) & (135.3, 135.6) & (148.8, 149.2)\\
$\sum_i \sum_{j \neq i} |b_{ij}|H(-b_{ij}b_{ji})$ & 151.6 & (139.2, 139.6) & (170.2, 170.6) & (154.3, 154.7)
\end{tabular}
\label{table:CSDSS}
\end{table*}

\begin{figure*}[t!]
\centering
\centerline{\includegraphics[scale=0.45]{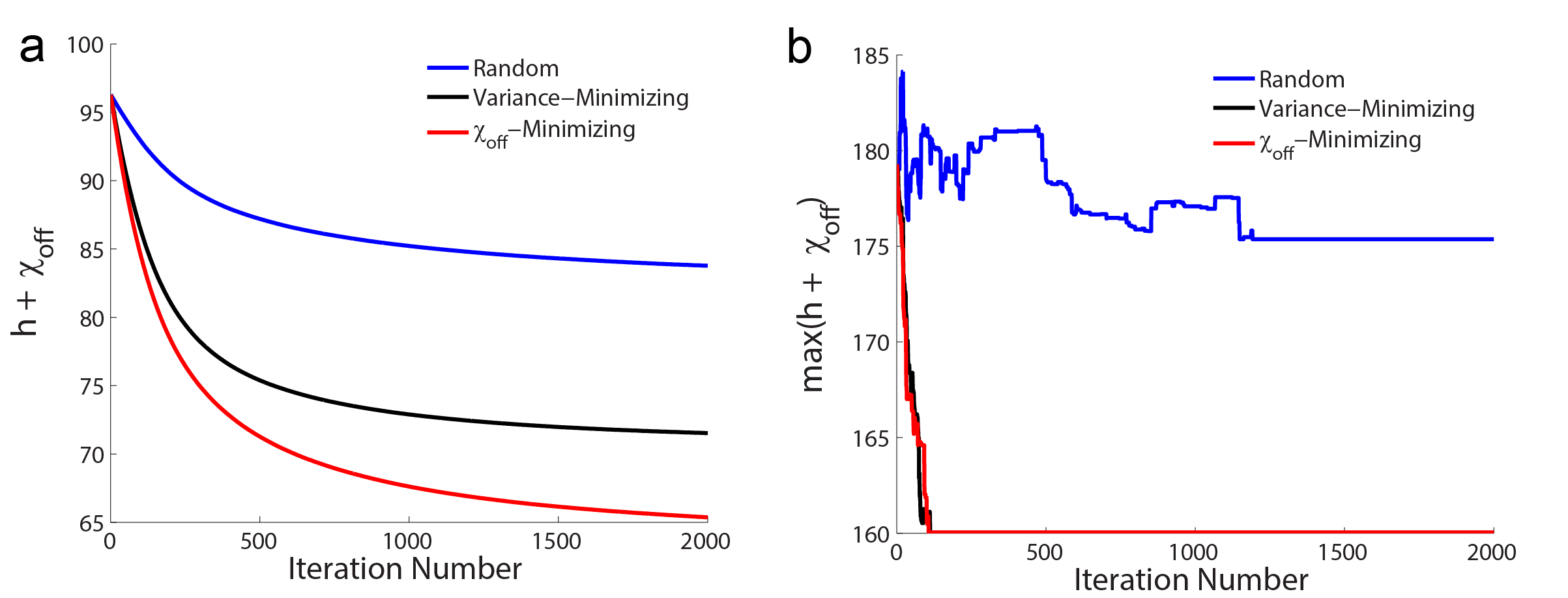}}
\caption{Plots of (a) the average $h+\chi_{\text{off}}$ and (b) the maximum $h+\chi_{\text{off}}$ against iteration number for over 50,000 implementations of the stabilization algorithm iteration number using the three different stabilization strategies. The initial random matrices are generated as per the main text. Standard error of the mean is on the order of $10^{-2}$ for subplot (a). } 
\label{fig:hc}
\end{figure*}

\clearpage

\subsection*{Supplementary results for the ecological model of constrained interaction strengths}
\subsubsection*{Analytical calculations of the $C$ vs $S$ relationship}

\begin{defi} \label{def:Initialization}
$C$ is defined as
\begin{align}
C(\mathbf{W_1},\mathbf{W_2}) = m^{-1} \sum_{k=1}^{m} \sum_{l=1}^{m} w_1^{kl} w_2^{lk} = m^{-1} \Tr(\bf{W}_1 \bf{W}_2) , 
\end{align}
where $k$ and $l$ are indices such that $k,l \in \{1 ,\dots m\}$ and $m$ is a positive integer i.e. $m\in \mathbb{N}$. The weights $w_1^{kl}$ and $w_2^{lk}$ for all $k$ and $l$ satisfy the weight constraints $\sum_k w_1^{kl} = 1$, $\sum_l w_1^{kl}=1$, $\sum_l w_2^{lk}=1$, $\sum_k w_2^{lk}=1$, $0 \leq w_1^{kl} \leq 1$, and $0 \leq  w_2^{lk} \leq 1$. Additionally, $w_1^{kl}$ is a matrix element of $\bf{W}_1$ with row index $k$ and column index $l$. Similarly, $w_2^{lk}$ is a matrix element of $\bf{W}_2$ with row index $l$ and column index $k$. 
\end{defi}

\begin{defi}
We define the measure of interdependency $S$ as 
\begin{align}
S(\mathbf{W_1}, \mathbf{W_2}) &= -\sum_{k} \sum_{l} \left(w_1^{kl}\right)^2 + \left(w_2^{lk}\right)^2 \\
&= -\Tr \left( \mathbf{W_1{W_1}^T} + \mathbf{W_2{W_2}^T} \right)
\end{align}
which is the negative sum of all the squared matrix elements of $W_1$ and $W_2$ of which $\mathbf{W_1}$ and $\mathbf{W_2}$ are elements respectively. 
\end{defi}

$S$ is a measure of diversity between the weight elements. A well known property of such a definition is that $S$ is maximized at $S=-2$ when all weight elements are equal at $1/m$ and minimized at $S=-2m$ when all weight elements are equal to one or zero. 

\begin{proposition} \label{prop:minCbound}
Let $\min(C)$ and $\max(C)$ denote respectively the minimum and maximum value of $C$ under a variation of the weight elements fulfilling the weight constraints. Then $\min(C)= 0$. Let $max(C)|_S$ and $min(C)|_S$ denote respectively the minimum and maximum value of $C$ under a variation of the weight elements fulfilling the weight constraints and under fixed $S$. Then, $min(C)|_{S=-2m}=0$. 
\end{proposition}
Since each weight is non-negative, each summand of $C$ must also be non-negative. Hence, $C$ is non-negative. Additionally, for $C=0$, each summand of $C$ must be equal to zero. Such a situation is possible when every matrix element is either zero or one at $S=-2m$. Hence, $\min(C)|_{S=-2m}=0$. 

\begin{proposition} \label{prop:minCzero}
$\min (C)|_S$ is (a) $0$ for $-2m \leq S \leq -4$ when $m$ is even and (b) $0$ for  $-2m \leq S \leq -4m^2/(m^2-1)$ when $m$ is odd. 
\end{proposition}

\textit{Proof.} Let $\{\mathbf{W_1}, \mathbf{W_2}\}$ represent a matrix pair such that $C(\mathbf{W_1},\mathbf{W_2})=0$. In order for $\min(C)=0$, it is necessary that all summands of $C$ are equal to zero. Hence, the total number of zeros from both matrices should be at least $m^2$. Let $\max(S_0)$ be the maximum $S$ such that $C=0$. Also, let $\{\mathbf{\underline{W}_1}, \mathbf{\underline{W}_2}\}$ be a matrix pair such that $C(\mathbf{\underline{W}_1},\mathbf{\underline{W}_2})=0$ and $S(\mathbf{\underline{W}_1}, \mathbf{\underline{W}_2}) = \max(S_0)$. 

(a) When $m$ is even, suppose that all weight constraints, $\sum_k w_1^{kl}=1$, $\sum_l w_1^{kl}=1$, $\sum_l w_2^{lk}=1$ and $\sum_k w_2^{lk}=1$ for all $k$ and $l$, are replaced by a less restrictive constraint $\sum_{k,l} w_1^{kl} + w_2^{lk}=2m$. Then for $C=0$, $S$ is maximized if each summand of $C$ is the multiplication of zero and $2/m$. However, each summand of $C$ can still be a multiplication of zero and $2/m$ if we were to use the original constraints instead. This is because it is possible to arrange $m/2$ zeroes in each row and column of each matrix without violating any weight constraints. In this case, $\max (S_0) = -4$. By Proposition \ref{prop:minCbound}, the result stated is obtained. 

(b) We first only consider the row constraints $\sum_l w_1^{kl}=1$ and $\sum_k w_2^{lk}=1$ without the column constraints  $\sum_k w_1^{kl}=1$ and $\sum_l w_2^{lk}=1$ for all $k$ and $l$. When $m$ is odd, the non-zero elements of each row in a matrix must be the same value within each row to maximize $S$. However, because $m$ is odd, the non-zero elements cannot be $2/m$. Also, to maximize $S$, there must not be more than a total of $m^2$ zeros in both matrices since any summand that is a multiplication of two zeros can still have $S$ increased. This is done by increasing the number of non-zero elements of either row possessing any of the two zeros. Since there is exactly a total of $m^2$ zeros in both matrices, then there must be $m$ rows each containing $(m+1)/2$ zeros and $m$ rows each containing $(m-1)/2$ zeros because any row that has more than $(m+1)/2$ zeros results in one or more other rows with less than $(m-1)/2$ zeros so that $S$ is necessarily smaller. Lastly, it is possible to arrange $m$ rows each containing $(m+1)/2$ zeros in one matrix and $m$ rows each containing $(m-1)/2$ zeros in the other matrix without violating any weight constraints, including the column constraints. Therefore, $\max(S_0)=-4m^2/(m^2-1)$. By Proposition \ref{prop:minCbound}, the result stated is obtained. 


\begin{theorem} \label{theorem:lagrange}
$\max(C)|_S$ is 
\begin{align}
C = -\frac{1}{2}Sm^{-1} \quad\quad \text{for } -2m \leq S \leq -4. 
\end{align}
When $m$ is even, $\min(C)|_S$ is
\begin{align}
mC = \begin{cases}
0, & \quad \text{for } -2m \leq S \leq -4, \\
\frac{1}{2}S + 2, & \quad \text{for } -4 \leq S \leq -2. 
\end{cases}
\end{align}
When $m$ is odd, $\min(C)|_S$ is 
\begin{align}
mC = \begin{cases}
0, & \quad \text{for } -2m \leq S \leq -\frac{-4m^2}{m^2-1}, \\
\frac{m-1}{2(m+1)}S + 4 \frac{2-m^2}{2(m+1)^2}, & \quad \text{for } -\frac{4m^2}{m^2-1} \leq S \leq \frac{2m-4}{m^2-1}-4, \\
S \frac{1-m}{2(m+1)} - \frac{2\sqrt{-2m(m-1)(S+4m+Sm)}+8}{(m+1)^2}-2+\frac{10}{m+1}, & \quad \text{for } \frac{2m-4}{m^2-1}-4 \leq S \leq -2\frac{2m-1}{m}, \\
\frac{1}{2}S + 2 & \quad \text{for } -2\frac{2m-1}{m} \leq S \leq -2. 
\end{cases}
\end{align}
\end{theorem}

The extrema may be found by using the method of Lagrange multipliers. In this case, the function to maximize/minimize is $C'= \sum_k \sum_l w_1^{kl}w_2^{lk}$, where $C'=mC$. The constraints are $\forall k: \sum_l w_1^{kl}=1$, $\forall l: \sum_k w_2^{lk}=1$, and $S=- \sum_k\sum_l{(w_1^{kl})}^2+{(w_2^{lk})}^2 $. The constraints $\forall l: \sum_k w_1^{kl}$ and $\forall k: \sum_l w_2^{lk}$ are not used, but we will show that the the solutions obtained from the simplified Lagrange multiplier problem can satisfy these two constraints that were left out. At the extrema, the gradient of $C'$ along a variable $w_1^{kl}$ is parallel to the gradient of the constraints along $w_1^{kl}$
\begin{align} \label{eq:Lagrange1}
w_2^{lk} = \zeta_k - 2 \rho w_1^{kl} 
\end{align}
Similarly, the gradient of $C'$ along a variable $w_2^{lk}$ is parallel to the gradient of the constraints along $w_2^{lk}$
\begin{align} \label{eq:Lagrange2}
w_1^{kl} &= \xi_l - 2\rho w_2^{lk} 
\end{align}
Here, $\xi_l$, $\zeta_k$, and $\rho$ are Lagrange multipliers. There are thus $2m^2$ equations corresponding to the $2m^2$ variables. Together with the constraints, there are $2m^2+2m+1$ equations corresponding to $2m^2$ variables and $2m+1$ Lagrange multipliers. We combine Equations \ref{eq:Lagrange1} and \ref{eq:Lagrange2} to obtain the following expressions for $w_1^{kl}$ and $w_2^{lk}$
\begin{align} \label{eq:Lagrangew2w1}
w_1^{kl} &= \frac{\xi_l - 2\rho \zeta_k }{1-4\rho^2} \\ \label{eq:Lagrangew1w2}
w_2^{lk} &= \frac{\zeta_k - 2\rho \xi_l}{1-4\rho^2}. 
\end{align}
These expressions give us $w_1^{kl}$ and $w_2^{lk}$ provided $\rho \neq \pm1/2$. Applying the constraint $\sum_l w_1^{kl} = 1$ to Equation \ref{eq:Lagrangew2w1} results in
\begin{align}
\sum_l \xi_l - 2m\rho \zeta_k = 1-4\rho^2
\end{align}
Since this equation must hold for any $k$, this implies that $\forall k: \zeta_k = \zeta$. Similarly, applying the constraint $\sum_k w_2^{lk}=1$ to Equation \ref{eq:Lagrangew1w2}, we find that $\forall l: \xi_l = \xi$. By Equations \ref{eq:Lagrangew2w1} and \ref{eq:Lagrangew1w2}, the weight elements in matrices $\mathbf{W_1}$ and $\mathbf{W_2}$ must be the same. However, this result is only valid when $S=-2$. When $S\neq -2$, then this result cannot hold. Hence when $S\neq -2$, then $\rho = \pm 1/2$. 

When $\rho = -1/2$, then summing all elements in $\mathbf{W_1}$ and $\mathbf{W_2}$ each using Equations \ref{eq:Lagrange1} and \ref{eq:Lagrange2} implies that $\forall k,l: \zeta_k=\xi_l=0$ since the elements in each matrix sum to $m$. Therefore, $\forall k,l: w_1^{kl}=w_2^{lk}$ and $S=-2C'$. This solution is the maximum as it does not preclude the possibility that the weights are non-negative within the domain of $S$. Additionally, since $\mathbf{W_1}=\mathbf{{W_2}^T}$, then all weights in each column of $\mathbf{W_1}$ and $\mathbf{W_2}$ must sum to one. 

When $\rho=1/2$, we see that $\forall k,l: \zeta_k=\xi_l=\zeta$ from Equation \ref{eq:Lagrangew2w1}. Summing all elements in $\mathbf{W_1}$ and $\mathbf{W_2}$ gives us $\zeta=2/m$, $\forall k,l: w_1^{kl}=2/m - w_2^{lk}$ and $S=2C'-4$. The solution $S=2C'-4$ is the minimum within the domain of $-4\leq S \leq -2$ if $m$ is even because the minimum does not preclude the possibility that the weights are non-negative within the domain of $-4\leq S\leq -2$. Additionally, since $\mathbf{W_1}=(2/m)\mathbf{J} - \mathbf{{W_2}^T}$, where $\mathbf{J}$ is an $m \times m$ matrix of ones, then each column of $\mathbf{W_1}$ and $\mathbf{W_2}$ must sum to one. 

If $m$ is odd, then $S=2C'-4$ is the minimum from $-2(2m-1)/m \leq S \leq -2$. The distribution at $S=-2(2m-1)/m$ corresponds to the situation where each row and each column of each matrix contains $(m-1)/2$ elements of zeros, $(m-1)/2$ elements of $2/m$ and an element of $1/m$. If $w_1^{kl} = 2/m$, then $w_2^{lk}=0$. If $w_1^{kl}=1/m$, then $w_2^{lk}=1/m$. This distribution represents the smallest $S$ attainable under the constraints defined because it is not possible to decrease $S$ further without violating the solution i.e. $w_1^{kl}=2/m-w_2^{lk}$, or having negative weights. We then reformulate the Lagrange multiplier problem for $-4m^2/(m^2-1)\leq S \leq -2(2m-1)/m$ by keeping the zeroes from $S=-2(2m-1)/m$ fixed within the domain of $-4m^2/(m^2-1)\leq S \leq -2(2m-1)/m$. 

If $w_2^{lk}=0$ and $w_1^{kl}$ is variable, then by Equation \ref{eq:Lagrange1}, 
\begin{align} \label{eq:Lagrangew1w20}
w_1^{kl} = \frac{\zeta_k}{2 \rho}. 
\end{align}
Similarly, if $w_1^{kl}=0$ and $w_2^{lk}$ is variable, then
\begin{align} \label{eq:Lagrangew1w21}
w_2^{lk} = \frac{\xi_l}{2 \rho}. 
\end{align}
If both $w_1^{kl}$ and $w_2^{lk}$ are variable, then $w_1^{kl}$ and $w_2^{lk}$ are given by Equations \ref{eq:Lagrangew2w1} and \ref{eq:Lagrangew1w2} respectively provided $\rho \neq \pm 1/2$. 

When $\rho = -1/2$, then $\zeta_k = - \xi_l$ if $w_1^{kl}$ and $w_2^{lk}$ are variable by Equation \ref{eq:Lagrangew2w1}. Given Equations \ref{eq:Lagrangew1w20} and \ref{eq:Lagrangew1w21}, this implies that $\forall k,l: \zeta_k=\xi_l=0$. Hence if $w_1^{kl}$ and $w_2^{lk}$ are both variable, then from Equation \ref{eq:Lagrangew2w1}, $w_1^{kl}=w_2^{lk}=1$ since the weights in each row sum to one. Therefore, we are not interested in this solution. 

With $\rho = 1/2$, then $\zeta_k=\xi_l$ if $w_1^{kl}$ and $w_2^{lk}$ are variable by Equation \ref{eq:Lagrangew2w1}. Then, summing over all $k$ for $w_1^{kl}$ is equal to summing over all $l$ for $w_2^{lk}$ which gives us $w_1^{kl}=w_2^{lk}=\zeta_k/2$. Since the sum over all $k$ for $w_1^{kl}$ is equal to one, then $\zeta_k=2/m$. Therefore, we are also not interested in this solution. 

When $\rho \neq \pm 1/2$, we may apply the weight constraints $\sum_k w_1^{kl}$ and $\sum_l w_2^{lk}$ on Equations \ref{eq:Lagrangew2w1}, \ref{eq:Lagrangew1w2}, \ref{eq:Lagrangew1w20} and \ref{eq:Lagrangew1w21} to find that
\begin{align} \label{eq:Brackets}
\zeta_k \left( \frac{m-1}{4\rho} - \frac{1}{1 - 2 \rho} \right) = \xi_l \left( \frac{m-1}{4\rho} - \frac{1}{1 - 2 \rho} \right), 
\end{align}
where $k$ and $l$ are such that $w_1^{kl}$ and $w_2^{lk}$ are variable. If the term in the brackets is not 0, then $\zeta_k = \xi_l$. We may then apply the weight constraint $\sum_k w_1^{kl}$ on Equations \ref{eq:Lagrangew2w1} and \ref{eq:Lagrangew1w20} again to obtain
\begin{align}
\zeta_k\left( \frac{m-1}{2} + \frac{1}{1+2\rho} \right) = 1. 
\end{align}
We can conduct the same analysis for Equations \ref{eq:Lagrangew1w2} and \ref{eq:Lagrangew1w21} to find $\xi_l$. Hence, it follows that $\zeta_k=\xi_l=\zeta$, 
\begin{align} \label{eq:xi1}
\zeta &= \frac{4\rho(1 + 2 \rho ) }{2\rho + m - 1 +2 \rho m}, \\ \label{eq:C1}
C' &= m\frac{\zeta^2}{(1+2\rho)^2}, 
\end{align}
and
\begin{align} \label{eq:S1}
S = -2C' - \frac{m (m-1) \zeta^2}{4\rho^2}. 
\end{align}
Combining Equations \ref{eq:xi1}, \ref{eq:C1} and \ref{eq:S1} results in two solutions for $S$, 
\begin{align} \label{eq:sc1a}
S=-2\frac{C' + 2m + C'm + 4m\sqrt{C'/m}}{m-1} \\ \label{eq:sc1b}
S=-2\frac{C' + 2m + C'm - 4m\sqrt{C'/m}}{m-1}
\end{align}
and for $dS/dC'$ respectively, 
\begin{align} \label{eq:dsdc1a}
\frac{dS}{dC'} &= -2-4\frac{\sqrt{\frac{m}{C'}}+1}{m-1} \\ \label{eq:dsdc1b}
\frac{dS}{dC'} &= -2+4\frac{\sqrt{\frac{m}{C'}}-1}{m-1}. 
\end{align}
We are not interested in Equation \ref{eq:sc1a} and \ref{eq:dsdc1a} because $dS/dC'$ is negative. When $S=-2(2m-1)/m$, then $C'=1/m$ at the minimum and $dS/dC'=2$. As $S$ is decreased further, $dS/dC'>2$ by Equation \ref{eq:dsdc1b}. For this solution, the weight elements of each column in $\mathbf{W_1}$ and $\mathbf{W_2}$ also sum to one. We are now left with the case when the term in the brackets in Equation \ref{eq:Brackets} is 0. 

When this term is 0, then 
\begin{align}
\rho= \frac{m-1}{2(m+1)}. 
\end{align}
For any $k$ where $w_1^{kl}$ and $w_2^{lk}$ are both variable, let $l=r(k)$. Therefore, $r(k)$ is a bijective function. Then $C'$, $S$ and the weight constraint are respectively, 
\begin{align} \label{eq:C2}
C' &= \mathlarger{\mathlarger{\sum}}_k \left(\frac{\xi_{r(k)}-2\rho\zeta_k}{1-4\rho^2}\right) \left(\frac{\zeta_k-2\rho\xi_{r(k)}}{1-4\rho^2}\right), \\ \label{eq:S2}
-S &= \mathlarger{\mathlarger{\sum}}_k \left(\frac{\xi_{r(k)}-2\rho \zeta_k}{1-4\rho^2}\right)^2 + \frac{m-1}{2}\left(\frac{\zeta_k}{2 \rho}\right)^2 +   \mathlarger{\mathlarger{\sum}}_l \left(\frac{\zeta_{r^{-1}(l)}-2\rho \xi_l}{1-4\rho^2}\right)^2 + \frac{m-1}{2}\left(\frac{\xi_l}{2 \rho}\right)^2, 
\end{align}
and
\begin{align} \label{eq:wcon2}
\frac{\xi_{r(k)} - 2\rho \zeta_k}{1-4\rho^2} + \frac{\zeta_k(m-1)}{4\rho} = 1. 
\end{align}
Combining Equations \ref{eq:C2}, \ref{eq:S2} and \ref{eq:wcon2}, we obtain for $S$ and $dS/dC'$, 
\begin{align} \label{eq:sc2}
S = 2C' + \frac{4C'+4C'm-4}{m^2-1} -4
\end{align}
and 
\begin{align} \label{eq:dsdc2}
\frac{dS}{dC'} &= \frac{4}{m-1} + 2. 
\end{align}
The columns in $\mathbf{W_1}$ and $\mathbf{W_2}$ do not necessarily sum to one for this solution. However, this solution can accommodate such a constraint by demanding that $\forall k: \zeta_k = \zeta$ and $\forall l: \xi_l = \xi$. Hence, Equation \ref{eq:sc1b} or \ref{eq:sc2}, depending on which has a smaller $C'$, gives the minimum for $-4m^2/(m^2-1)\leq S \leq -2(2m-1)/m$ since the solutions do not preclude the possibility that the weights are non-negative for this domain of $S$. Specifically, for $(2m-4)/(m^2-1)-4 \leq S \leq -2(2m-1)/m$, the minimum is given by Equation \ref{eq:sc1b} while for $-4m^2/(m^2-1) \leq S \leq (2m-4)/(m^2-1)-4$, the minimum is given by Equation \ref{eq:sc2}. Along with Proposition \ref{prop:minCzero}, the results stated in the theorem are obtained. 

\begin{theorem}
The symmetric correlation is bounded $0\leq C \leq 1$. 
\end{theorem}
It follows from Theorem \ref{theorem:lagrange} that $C$ is bounded $0\leq C \leq 1$. 

\subsubsection*{A remark about the general case where $W_1$ and $W_2$ are not square matrices}
When $y$ and $z$ are of different sizes i.e. $|y| = p$ and $|z| = q$, where $p$ and $q$ are positive integers and $p>q$, then $\mathbf{W_1}$ becomes a $p \times q$ matrix and $\mathbf{W_2}$ becomes a $q \times p$ matrix. In this case, we define $C(\mathbf{W_1},\mathbf{W_2}) = q^{-1} \Tr(\mathbf{W_1}\mathbf{W_2})$. The relationship between $\max(C)|_S$ and $S$, and $\min(C)|_S$ and $S$ does not appear to be trivial to derive for the more general case when $y$ and $z$ are of different sizes. However, we can still expect trophic relationships to be more adversely affected than mutualistic and competitive relationships because at minimum interdependence diversity , $\max(C)|_{S=-(p+q)}=1$ and $\min(C)|_{S=-(p+q)}=0$ whereas at maximum interdependence diversity ($S=-(q^2+p^2)/pq$), we find that $C$ only has one possible value at $C=q^{-1}$. 
\bibliographystyle{unsrt}
\bibliography{References}

\end{document}